\documentclass[useAMS,usenatbib,usegraphicx]{mn2e}

\usepackage{amsmath,amssymb,hyperref,color,soul,nicefrac}
\usepackage{txfonts}

\newcommand{\sgr}{\mbox{Sgr~A$^\ast$}}

\title[Scattering Properties of the Milky Way]{Milky Way scattering properties and intrinsic sizes of active galactic nuclei cores probed by very long baseline interferometry surveys of compact extragalactic radio sources}
\author[A. B. Pushkarev and Y. Y. Kovalev]{A. B. Pushkarev$^{1,2}$\thanks{E-mail:
pushkarev.alexander@gmail.com (ABP)} and Y. Y. Kovalev$^{2,3}$\\
$^{1}$Crimean Astrophysical Observatory, Nauchny 298688, Crimea, Russia\\
$^{2}$Astro Space Center of Lebedev Physical Institute, Profsoyuznaya 84/32, Moscow 117997, Russia\\
$^{3}$Max-Planck-Institut f\"ur Radioastronomie, Auf dem H\"ugel 69, 53121 Bonn, Germany}
\begin{document}

\date{Accepted 2015 July 08. Received 2015 July 02; in original form 2015 April 30}

\pagerange{\pageref{firstpage}--\pageref{lastpage}} \pubyear{2015}

\maketitle

\label{firstpage}

\begin{abstract}
We have measured the angular sizes of radio cores of active galactic nuclei (AGN) and analyzed 
their sky distributions and frequency dependencies to study synchrotron opacity in AGN jets and 
the strength of angular broadening in the interstellar medium. We have used archival very 
long baseline interferometry (VLBI) data of more than 3000 compact extragalactic radio sources 
observed at frequencies, $\nu$, from 2 to 43~GHz to measure the observed angular size of VLBI 
cores. We have found a significant increase in the angular sizes of the extragalactic sources 
seen through the Galactic plane ($|b|\la10\degr$) at 2, 5 and 8 GHz, about \nicefrac{1}{3} of which show 
significant scattering. These sources are mainly detected in directions to the Galactic bar, 
the Cygnus region, and a region with galactic longitudes $220\degr\la l\la260\degr$ (the 
Fitzgerald window). The strength of interstellar scattering of the AGNs is found to correlate 
with the Galactic H$\alpha$ intensity, free-electron density, and Galactic rotation measure. 
The dependence of scattering strengths on source redshift is insignificant, suggesting that 
the dominant scattering screens are located in our Galaxy. The observed angular size of 
\sgr\ is found to be the largest among thousands of AGN observed over the sky; we discuss
possible reasons of this strange result. Excluding extragalactic radio sources with significant 
scattering, we find that angular size of opaque cores in AGN scales typically as $\nu^{-1}$ 
confirming predictions of a conical synchrotron jet model with equipartition. \end{abstract}


\section{Introduction}

Radio waves emitted by a compact background radio source are influenced by
propagation effects whenever passing through an ionized medium containing
free-electron density fluctuations $\Delta n_e/n_e$. Diffraction phenomena
associated with scattering results in the angular broadening of a compact
bright source, the scattered size of which scales as $\nu^{-k}$, where $k\sim2$
depending on the form of the spatial power spectrum of electron density turbulence
\citep[e.g.,][]{Goodman85,Cordes86}. If no scattering is present, the observed 
angular size of opaque cores (i.e., apparent jet base) in AGN is expected to 
scale approximately as $\nu^{-1}$  \citep[][see Sect.~\ref{s:jet_theory} for 
model assumptions]{BK79,Koenigl81}. Departures from this dependence are also 
possible and can be caused by pressure and density gradients in the jet or by 
external absorption from the surrounding medium 
\citep[][and references therein]{Lobanov98,Kovalev_cs,Sokolovsky_cs,Pushkarev_cs}.
This theoretical expectation has never been confirmed statistically by observations 
of large samples of AGN radio cores.

Based on theoretical models, intergalactic scattering is negligible and can be barely 
detected with either space VLBI or a low-frequency interferometer. In the most optimistic 
scenario, when a background radio source intersects a galaxy halo with overdensity 
$\sim$1000 relative to the mean baryon density of the Universe and the outer scale of 
turbulence $\sim$1~kpc, the scatter broadening contribution by the intergalactic medium 
would be about 0.1~mas at 2~GHz \citep{Koay15}. Therefore, the angular broadening is 
expected to be strongly dominated by interstellar scattering in the Galaxy.

After about two decades since the VLBA's commissioning in 1994 \citep{nap94}, 
thousands of active galactic nuclei have been observed to date with VLBA and 
global VLBI arrays at many frequency bands (see for details Sect.~\ref{s:data_meas}), 
thereby making possible the investigation of the scattering properties within and outside 
of the Galactic plane. Particularly useful are massive dual-band truly simultaneous 
2.3 and 8.4~GHz VLBI observations originally driven by astrometric and geophysical 
applications. We use these results to (i) investigate scattering properties of Milky 
Way, and (ii) measure scatter-free frequency dependence of core sizes in AGN jets 
and compare it with theoretical predictions.

\begin{figure}
\centering
\includegraphics[angle=-90,width=0.48\textwidth]{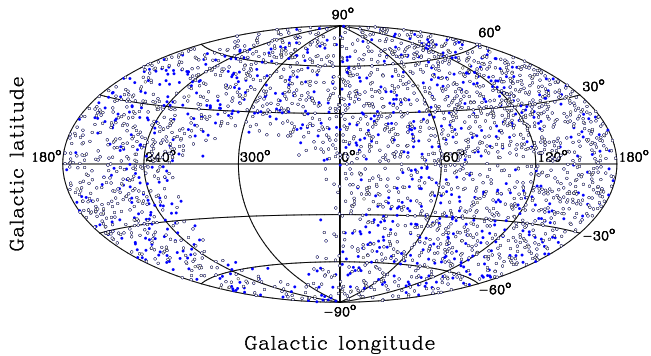}
\caption{
Sky distribution of 3019 RFC sources shown in Aitoff equal-area projection of the celestial
sphere in galactic coordinates and used in the analysis after applying the selection criteria.
Open dots represent 2327 sources observed simultaneously at 2 and 8~GHz. Blue filled dots
denote the other 692 objects observed at 8~GHz only. The sources shown here exclude 
those with low core S/N or unresolved cores.
\label{f:sky_distr}
}
\end{figure}

\section{Multi-frequency data and measurements of AGN core sizes}
\label{s:data_meas}

For the purposes of our analysis we made use of data from the Radio 
Fundamental Catalog (RFC\footnote{\url{http://astrogeo.org/rfc}}) that 
comprises all sources observed with the VLBI under absolute astrometry 
and geodesy programs from 1980 through 2014 as well as other major VLBI 
surveys, described below. As the observed AGN are highly variable objects, 
we need data taken simultaneously at different frequencies to analyze 
the effect of angular broadening. Therefore, we used the truly simultaneous, 
the most complete and deep survey data at S and X-bands centered at 2.3 
and 8.4~GHz, respectively. The frequencies are rounded to integer values 
later in the text for simplicity. These data were mostly obtained within 
the VLBA Calibrator Survey \citep[VCS;][]{vcs1,vcs2,vcs3,vcs4,vcs5,vcs6} 
and Research \& Development -- VLBA \citep[RDV;][]{RDV2009} observing 
sessions \citep[e.g.,][]{PK12, Piner12}. A total of 11607 epochs of 
observations were used to observe the 3778 sources with declinations above 
$\approx-45\degr$ at 2~GHz, while a total of 12142 epochs were used to 
observe 4096 sources distributed over the entire sky at 8~GHz. Additionally, 
we made some use of data at 5~GHz mainly coming from the VLBI Imaging and 
Polarimetry Survey 
\citep[VIPS;][]{Hel07,Petrov_VIPS}, and the VLBA Calibrator Survey (VCS-7, VCS-8; 
Petrov, in prep.);
at  8~GHz from the VLBA+GBT observing program for \textit{Fermi}-AGN associations \citep{Kovalev_Fermi11};
at 15~GHz from the VLBA 2\,cm Survey \citep{2cmPaperI,2cmPaperIII}, 
          the Monitoring Of Jets in Active galactic nuclei with VLBA Experiments \citep[MOJAVE;][]{MOJAVE_maps,lister13};
at 24~GHz from the VLBI Exploration of Radio Astrometry (VERA) and VLBA K-band \citep{Petrov_etal07,VLBA_K_band}
          and the ICRF at higher frequencies \citep{Charlot10_KQ};
at 43~GHz from the QCAL-1 43~GHz Calibrator Survey \citep{Petrov12_QCAL} and 
          the VLBA-BU Blazar Monitoring Program\footnote{\url{http://www.bu.edu/blazars/VLBAproject.html}}.

\begin{figure*}
\centering
\includegraphics[angle=-90,width=0.454\textwidth]{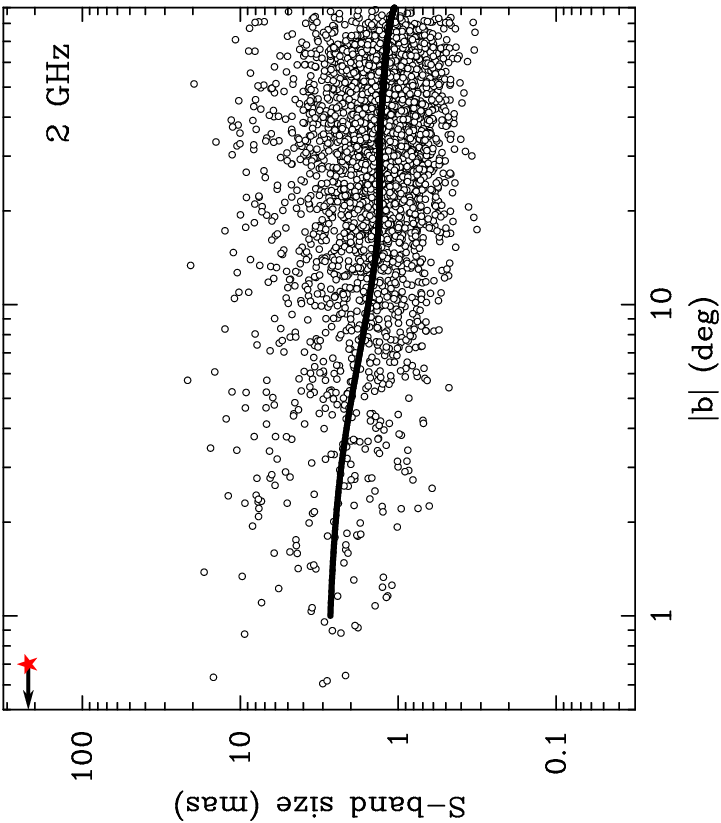}\hspace{0.3cm}
\includegraphics[angle=-90,width=0.454\textwidth]{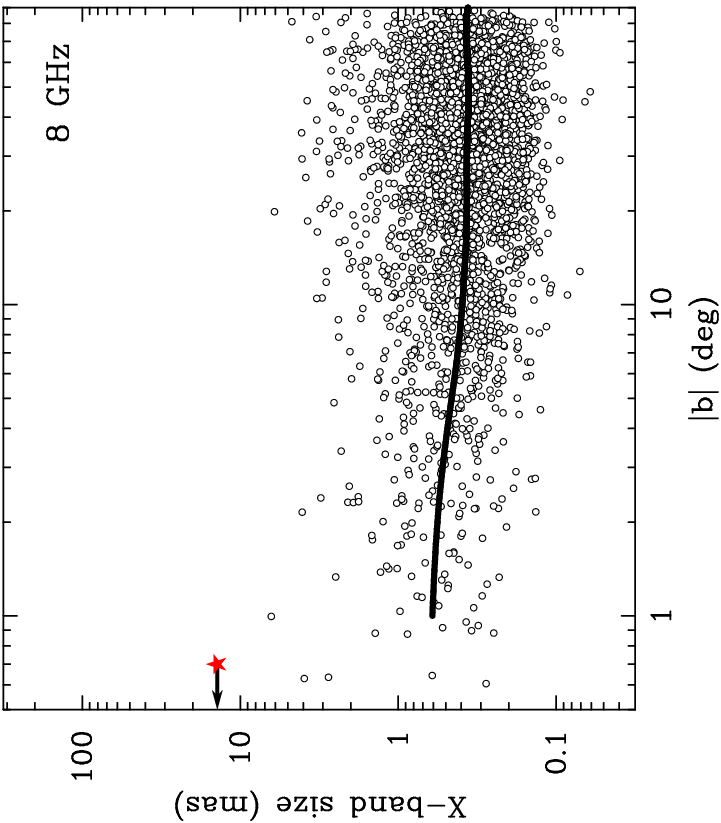}\vspace{0.3cm}
\includegraphics[angle=-90,width=0.454\textwidth]{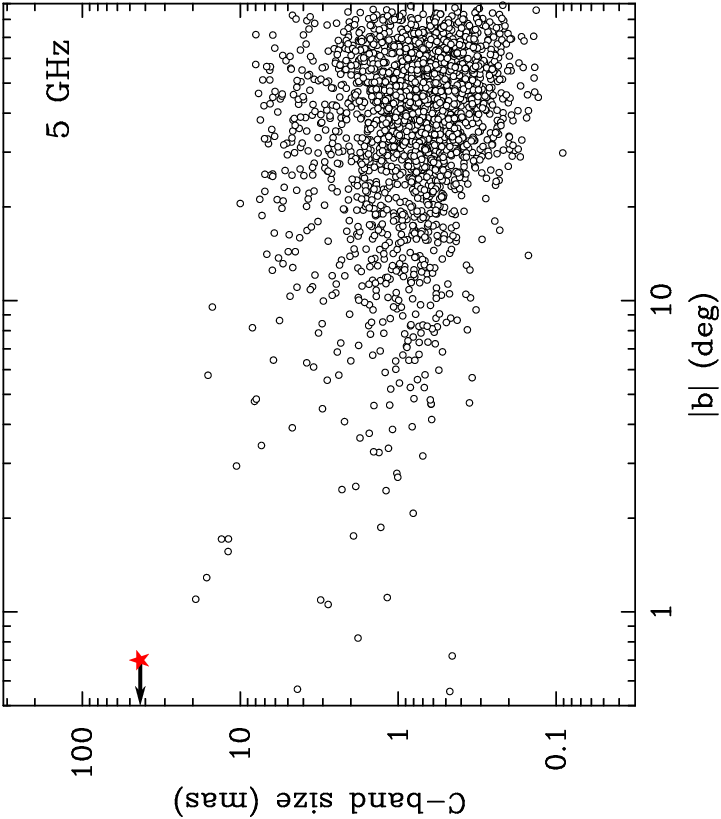}\hspace{0.3cm}
\includegraphics[angle=-90,width=0.454\textwidth]{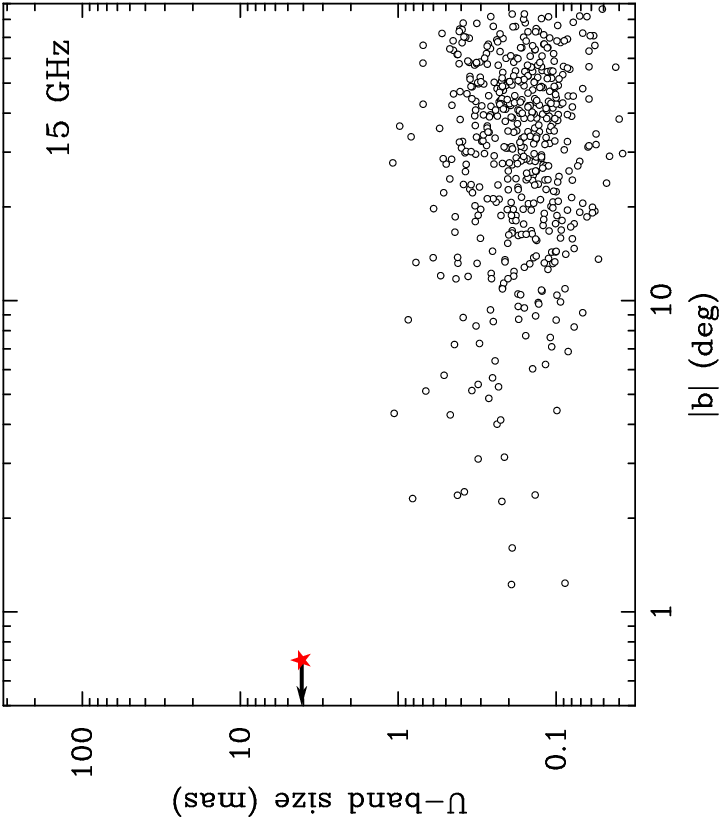}\vspace{0.3cm}
\includegraphics[angle=-90,width=0.454\textwidth]{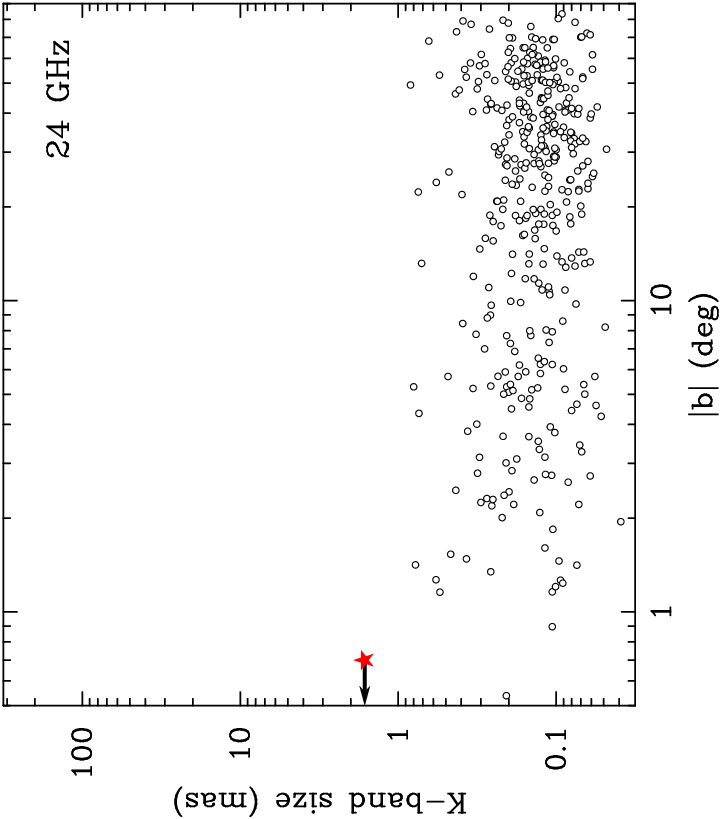}\hspace{0.3cm}
\includegraphics[angle=-90,width=0.454\textwidth]{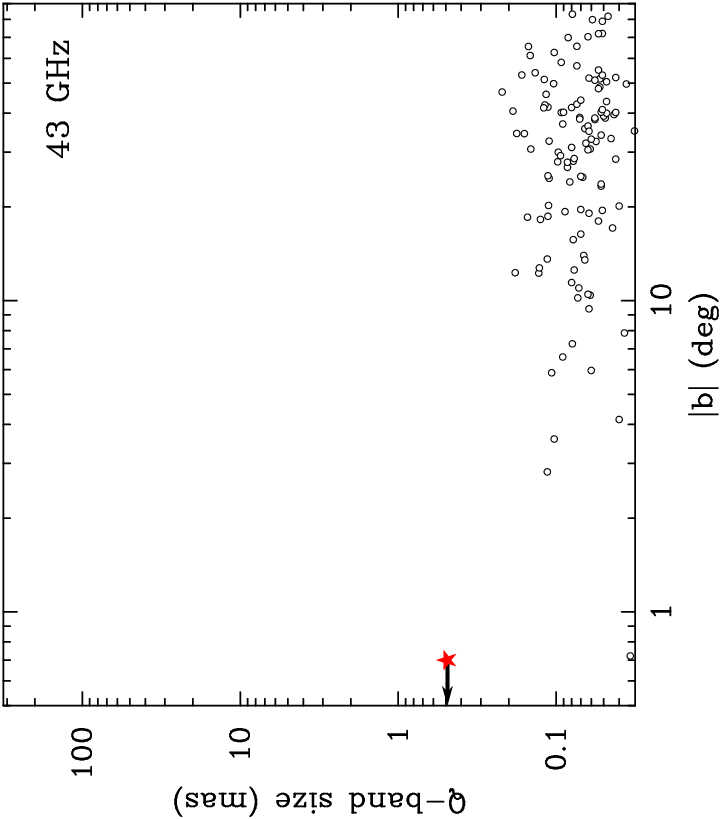}\vspace{0.3cm}
\caption{
Observed full width at half maximum (FWHM) angular size of VLBI core component of AGN at
2, 5, 8, 15, 24, and 43~GHz versus the absolute value of galactic latitude. Each point
represents a single source; a median is used for sources with size measurements at more
than one epoch. Thick lines represent the cubic spline interpolation and are shown for
the 2 and 8~GHz data only, characterized by the best completeness properties, most uniform
sky distribution and the largest number of sources. \sgr\ is shown by a star symbol.
Correlation statistics is presented in Table~\ref{t:size_vs_b}.
\label{f:size_vs_b}
}
\end{figure*}

Parsec-scale morphology of AGN probed by VLBI observations is typically
represented by a one-sided core-jet structure, with a compact core that 
dominates the total flux density, and a weaker outflow that quickly dims 
downstream. Therefore, a model of two circular Gaussian components has been 
fitted to the self-calibrated visibility data for all sources, with the most
 compact component assigned as the VLBI core following the approach by \citet{2cmPaperIV}.

\begin{table}
\caption{Kendall's $\tau$-test correlation statistics for the measured size of cores observed
         over the entire sky at different frequencies versus the absolute galactic latitude 
         shown in Fig.~\ref{f:size_vs_b}. Errors are given at 95\% level of significance.}
\label{t:size_vs_b}
\centering
\begin{tabular}{rcrc}
\hline
\noalign{\smallskip}
$\nu$, GHz & $\tau$ & $N$ & $p$ \\
   (1)     &    (2) & (3) & (4) \\
\hline\noalign{\smallskip}
 2.3 & $-0.136\pm0.012$ & 2888 & $5\times10^{-28}$    \\
 5.0 & $-0.135\pm0.013$ & 2108 & $2\times10^{-20}$    \\
 8.4 & $-0.051\pm0.011$ & 3019 & $3\times10^{-5}$     \\
15.4 & $-0.015\pm0.030$ &  513 & 0.61     \\
24.4 & $-0.026\pm0.033$ &  402 & 0.43     \\
43.1 & $-0.063\pm0.060$ &  119 & 0.31     \\
\hline
\end{tabular}

\smallskip
Note: $N$ is the number of sources, $p$ is the probability that the correlation occurred by chance.
\end{table}

\begin{table*}
\begin{minipage}{126mm}
\caption{Apparent angular sizes of VLBI cores of 4963 AGN determined by a modelfit. 
         The full table is available online at \url{ftp://arc.u-strasbg.fr/pub/cats/J/MNRAS/452/4274/}.}
\label{t:sizes_ind}
\begin{tabular}{c r r c c c c c c c}
\hline
\noalign{\smallskip}
             &        $b$  &      $l$  &       $z$ & $\theta_{2}$& $\theta_{5}$& $\theta_{8}$&$\theta_{15}$&$\theta_{24}$&$\theta_{43}$ \\
        Name &      (deg)  &    (deg)  &           &       (mas) &       (mas) &       (mas) &       (mas) &       (mas) &       (mas) \\
         (1) &        (2)  &      (3)  &       (4) &         (5) &         (6) &         (7) &         (8) &         (9) &        (10) \\
\hline\noalign{\smallskip}
J0000$-$3221 &   $-$77.752 &   357.466 &     1.275 &        0.98 &      \ldots &        0.45 &      \ldots &      \ldots &      \ldots \\
J0000$+$4054 &   $-$20.968 &   112.163 &    \ldots &        3.65 &      \ldots &        2.90 &      \ldots &      \ldots &      \ldots \\
J0001$+$1456 &   $-$46.228 &   104.539 &     0.399 &      \ldots &        1.16 &      \ldots &      \ldots &      \ldots &      \ldots \\
J0001$+$1914 &   $-$42.054 &   105.998 &     3.100 &        0.94 &      \ldots &        0.19 &      \ldots &      \ldots &      \ldots \\
J0001$+$4440 &   $-$17.292 &   113.181 &    \ldots &      \ldots &        0.46 &      \ldots &      \ldots &      \ldots &      \ldots \\
J0001$+$6051 &    $-$1.419 &   116.869 &    \ldots &        4.30 &      \ldots &        1.02 &      \ldots &      \ldots &      \ldots \\
J0002$-$2153 &   $-$77.642 &    52.802 &    \ldots &      \ldots &      \ldots &        0.24 &      \ldots &      \ldots &      \ldots \\
J0003$-$1547 &   $-$74.105 &    75.364 &     0.508 &      \ldots &      \ldots &        0.54 &      \ldots &      \ldots &      \ldots \\
J0003$-$1927 &   $-$76.560 &    63.669 &     2.000 &        1.52 &      \ldots &        0.94 &      \ldots &      \ldots &      \ldots \\
J0003$+$2129 &   $-$40.002 &   107.437 &     0.450 &      \ldots &      \ldots &        0.34 &        0.17 &      \ldots &      \ldots \\
\hline
\end{tabular}

\smallskip
Columns are as follows: (1) source name in J2000.0 notation; (2) Galactic 
latitude; (3) Galactic longitude; (4) redshift from the continuously updated 
OCARS database at {\url{http://www.gao.spb.ru/english/as/ac_vlbi/ocars.txt}}.
Reference to original redshift data can be found in this compilatory database,
description of which is given in \citet{Malkin08}; (5)--(10) FWHM of the fitted 
circular Gaussian at 2, 5, 8, 15, 24, and 43~GHz, respectively. If a source was 
observed over multiple epochs, the median of the observed angular size is given. 
This table is available in its entirety in a machine-readable form online.
A portion is shown here for guidance regarding its form and content.
\end{minipage}
\end{table*}

The structure modeling has been performed with the procedure {\it modelfit} 
in the Difmap package \citep{difmap} using an automated approach. Cores were 
found to be unresolved in a fraction of epochs for a fraction of sources, 
i.e.\ their fitted angular sizes were smaller than a corresponding resolution 
limit calculated following \citet{2cmPaperIV}. In rare cases the core components 
had a low ($<10$) signal-to-noise ratio. We excluded data for such epochs to get 
rid of a bias they may introduce. In cases where multi-epoch data can be used 
for the core size estimates, a single median value was used per source in the 
analysis. The resulting samples contain 2888 and 3019 sources at 2 and 8~GHz 
(Table~\ref{t:size_vs_b}), respectively, with measured sizes of the cores, of 
which 2327 sources have simultaneous observations at 2 and 8~GHz (Fig.~\ref{f:sky_distr}).

\section{Apparent core size inside and outside the Galactic plane}
\label{s:app_core_sizes}

In Fig.~\ref{f:size_vs_b} we plot the angular size $\theta$ of the cores as a 
function of the absolute value of galactic latitude $|b|$, as scattering effects 
are most essential for compact components. The general behavior shows that sources 
in the Galactic plane ($|b|\la10\degr$), comprising about 14\% of the samples at 2 
and 8~GHz, have on average larger apparent angular sizes relative to sources at 
higher galactic latitudes. In particular, at 8~GHz the spline values gradually 
decrease by a factor of $\sim$1.7 between $|b|\approx1\degr$ and $|b|\approx10\degr$ 
and then remain nearly constant at larger latitudes. At lower frequency, the 2~GHz 
spline shows similar behavior but with the size decrease present along the entire 
range of $|b|$. At both frequencies the non-parametric Kendall's $\tau$-test confirms 
a highly significant ($p_\mathrm{8\,GHz}=2\times10^{-6}$, $p_\mathrm{2\,GHz}=10^{-9}$) 
negative correlation between $\theta$ and $|b|$ within the Galactic plane. At 5~GHz 
the correlation is also highly significant, while at higher frequencies, 15, 24, and 
43~GHz scattering is much weaker with no significant correlation observed. The 
all-sky correlation statistics for the measured observed size of the VLBI core at 
different frequencies against the absolute value of the galactic latitude is summarized 
in Table~\ref{t:size_vs_b}, where we list observing frequency in column (1), 
ranked correlation coefficient in column (2), number of sources in column (3), and 
probability of correlation by chance in column (4). We estimated the uncertainties 
of the non-parametric Kendall's $\tau$ correlation coefficients listed in 
Tables~\ref{t:size_vs_b} and \ref{t:k_vs_rm_ne_ha} at a significance level of 95\%,
using a randomization technique based on (i) randomly selecting 80\% of a sample, 
(ii) calculating the $\tau$, (iii) repeating steps (i) and (ii) 2000 times and 
then constructing the confidence intervals. The VLBI core sizes measured at 
different frequencies in a range from 2 to 43~GHz for 4963 sources used in our 
analysis and presented in Fig.~\ref{f:size_vs_b} are listed in Table~\ref{t:sizes_ind}. 
We note that for sources with rich parsec-scale morphology the estimated core sizes 
might depend strongly on the model chosen to represent the structure.

\begin{figure}
\centering
\includegraphics[angle=-90,width=0.45\textwidth]{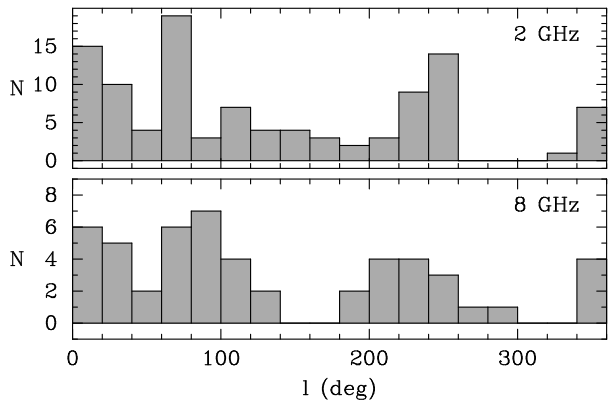}
\caption{
Histograms of the galactic longitude for the large-sized Galactic plane sources from
Fig.~\ref{f:size_vs_b} with angular size exceeding 5~mas at 2~GHz (top) and 1~mas at 8~GHz (bottom).
\label{f:size_vs_l}
}
\end{figure}

\begin{figure*}
\centering
\resizebox{0.48\hsize}{!}{\includegraphics[angle=-90]{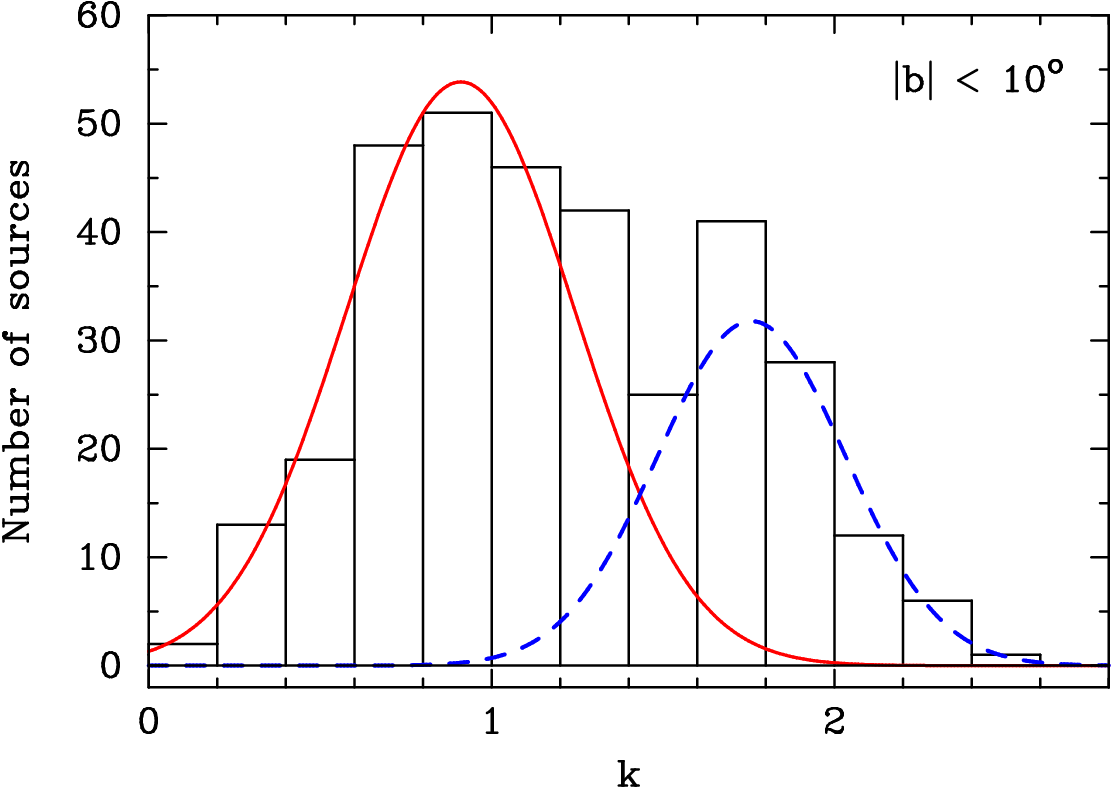}}\hspace{0.3cm}
\resizebox{0.48\hsize}{!}{\includegraphics[angle=-90]{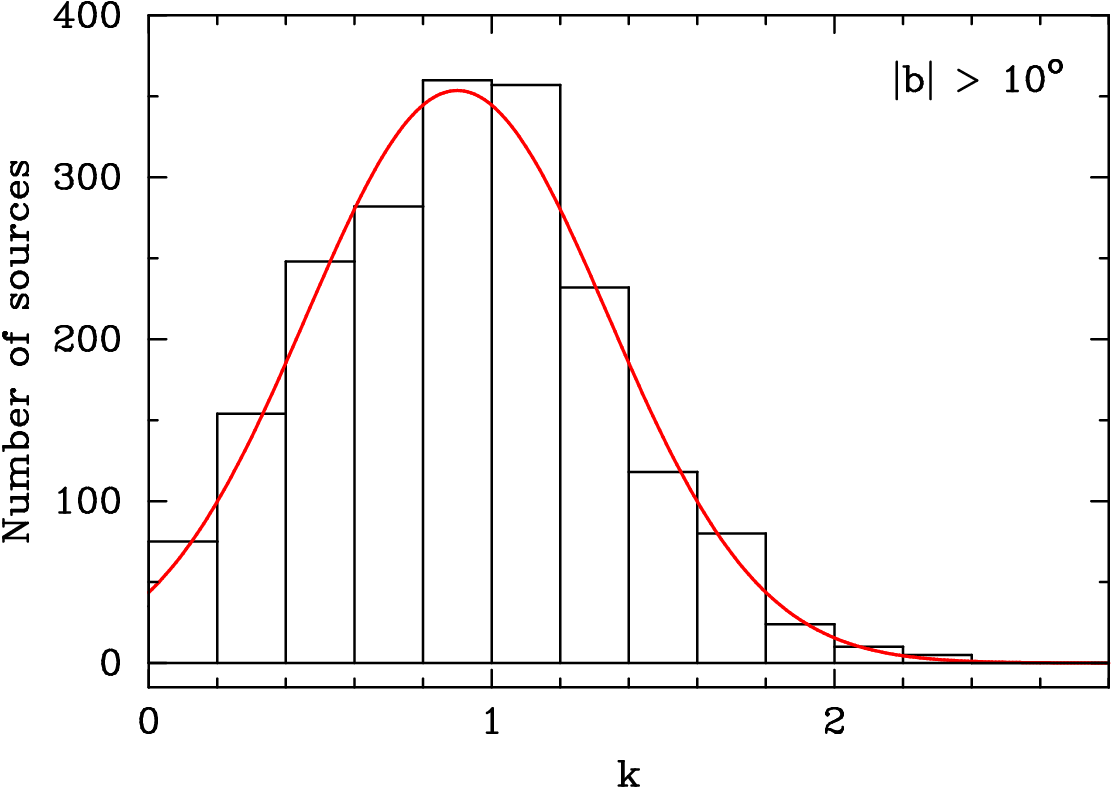}}
\caption{Histograms of the $k$-index in a size-frequency dependence $\theta\propto\nu^{-k}$ for 
the sources within the Galactic plane (left) and away from it (right). For sources with multi-epoch 
observations median values of $k$ are used. Two Gaussians are fitted to the left-panel distribution 
representing non-scattered ($\mu=0.91$) and scatter-broadened ($\mu=1.76$) sources. The right-panel 
distribution is fitted with a single Gaussian. Details on the fits are shown in Table~\ref{t:gaussians}. 
Bins with negative $k$ values are not shown, comprising about 0.3\% and 2.4\% of the sources 
with $k$ down to $-0.01$ and $-0.47$ at the low and high Galactic latitudes, respectively.}
\label{f:k_hists}
\end{figure*}

The sources with large angular sizes ($>1$~mas at 8~GHz and $>5$~mas at 2~GHz) within 
the Galactic plane are non-uniformly distributed with the galactic longitude $l$ 
(Fig.~\ref{f:size_vs_l}). The distribution peaks in the regions $340\degr\la l\la20\degr$ 
representing the direction to the Galactic center (GC) and bar, $65\degr\la l\la90\degr$ 
towards the Cygnus region, and $220\degr\la l\la260\degr$ towards star-forming regions in 
the Perseus and Local (Orion) arms \citep{Vazquez08}, and the Vela supernova remnant.
Screens with significant scattering strength are expected in these regions. The 
patchy distributions of scatter-broadened compact extragalactic sources will be analyzed 
and presented by us elsewhere. We also note the lack of the broadened sources along the 
direction to the Galactic anticenter ($l\sim180\degr$), where less interstellar 
scattering is expected within the Galactic plane. The gap of sources around $l\sim300\degr$ 
is due to the absence of the southernmost AGN in available VLBI surveys, especially at 
2~GHz, as can be seen from the sky distribution in Fig.~\ref{f:sky_distr}.

Since the Sun is about $26\pm3$~pc away from the Galactic plane as inferred from classical
Cepheids \citep{Majaess09}, we tested for possible asymmetry in the angular size distributions
for the sources with negative and positive galactic latitude within the Galactic plane. 
The latter population shows an $\sim$8\% excess in quantity (Fig.~\ref{f:sky_distr}) but 
no significant difference in the angular size at 2 and 8~GHz is found by a K-S test. Mean 
values also agree within the errors, with $3.0\pm0.3$~mas ($0\degr<b<10\degr$) and 
$2.6\pm0.3$ ($-10\degr<b<0\degr$) at 2~GHz and $0.60\pm0.06$~mas ($0\degr<b<10\degr$) 
and $0.63\pm0.08$ ($-10\degr<b<0\degr$) at 8~GHz. Additionally, no significant difference 
between $k$-index distributions for these groups of sources is found by a K-S test 
(see more details in Sect.~\ref{s:freq}).

\section{Separating intrinsic and external effects: frequency dependence of AGN core size}
\label{s:freq}

We compiled a sample of 2327 sources that have been observed at 2 and 8~GHz simultaneously 
and for which the $k$-index was successfully estimated for the size-frequency dependence 
$\theta\propto\nu^{-k}$ for each object using the corresponding measured VLBI core sizes. 
If a source had more than one epoch of observations, the median value of $k$ was derived. 
We show the resulting distribution of the $k$-index with two histograms: for 335 sources within 
the Galactic plane (Fig.~\ref{f:k_hists}, left, $|b|<10\degr$) and 1992 sources at $|b|>10\degr$ 
(Fig.~\ref{f:k_hists}, right). The $k$-index distribution for the sources at low galactic 
latitudes was fitted with two Gaussians the best-fit peaks of which are 0.91 and 1.76. 
The lower and higher peaked Gaussians represent two groups of sources, non-scattered and 
scattered by intervening screens, respectively. The $k$-index distribution for the sources 
out of the Galactic plane (Fig.~\ref{f:k_hists}, right) shows that most of them are not 
affected significantly by angular broadening. The distribution is fitted by a single Gaussian 
with the best-fit peak at 0.90, which is consistent with the left Gaussian for the Galactic 
plane AGN. Since the first peak is observed for both distributions (Table~\ref{t:gaussians}) 
while the second peak is visible for the Galactic plane only, this confirms that the latter 
has an origin that is extrinsic to the AGN.

\begin{table}
\caption{Parameters of the Gaussians fitted to the $k$-index distributions in Fig.~\ref{f:k_hists}.}
\label{t:gaussians}
\centering
\begin{tabular}{cccc}
\hline
\noalign{\smallskip}
$|b|$   & $\lambda$ & $\mu$ & $\sigma$ \\
   (1)  &    (2)    &  (3)  &   (4)    \\
\hline\noalign{\smallskip}
$<10\degr$ & 0.67 & 0.91 & 0.33 \\
           & 0.33 & 1.76 & 0.28 \\\cline{2-4}\noalign{\smallskip}
$>10\degr$ & 1.00 & 0.90 & 0.44 \\
\hline
\end{tabular}

\smallskip
Note: the presented parameters are found from the following Gaussian fit: 
$\lambda\exp\left(\frac{(k-\mu)^2}{2\sigma^2}\right)/(\sigma\sqrt{2\pi})$.

\end{table}

We note that only about \nicefrac{1}{3} of the sources within $|b|<10\degr$ shows
angular scatter-broadening external to the source. As the typical value of $k$ for
the scattered sources is less than the expected $k\approx2$, this may indicate that
the angular size of some turbulent eddies are comparable to that of the VLBI cores,
and refractive effects may also play a role in scattering \citep[e.g.,][]{Cordes86,Pushkarev13}.
Alternatively, it can be partly caused by variability of scattering properties of intervening
screen on the line of sight to a source \citep[e.g.,][]{Koay11_J1819,Pushkarev13,deBruyn15}.
If a scatter-broadened source has more than one epoch of observations and a scattering
screen either changes its characteristics or moves away, the median $k$ would decrease.
In total, about 30\% of the analyzed sources with $|b|<10\degr$ had two or more epochs of
observations, making scattering variability potentially responsible for the lower $k$. To
test this scenario, we repeated the analysis constructing a $k$-index distribution using (i)
the first epoch only for each source, (ii) the last epoch only. In all three cases, the
obtained distributions and fitted Gaussians are similar indicating that the slightly 
lower value of $k$ cannot be explained by a net movement of screens away from the line-of-sight.
To confirm that significant variability of scattering is a rare phenomenon on a time scale 
of few years, we analyzed distribution of $(k_2 - k_1)$ for each of the 108 sources 
within the Galactic plane that had more than one epoch, where $k_2$ is the value of $k$ in 
the last epoch and $k_1$ is the value of $k$ in the first epoch. Indeed, we obtained a narrow 
distribution peaking at 0, with a median value 0.01 and a standard deviation 0.55. 

Another possible reason we observe the lower values of $k$ in scatter broadened sources is 
that the scattering effects do not dominate. Since the observed angular size 
$\theta_\mathrm{obs} = (\theta^2_\mathrm{int} + \theta^2_\mathrm{scat})^{1/2}$,
where $\theta_\mathrm{int}$ is the intrinsic source size, and $\theta_\mathrm{scat}$ is 
the scatter broadening, the observed angular size will scale with $\nu^{-2}$ only if 
$\theta_\mathrm{scat} \gg \theta_\mathrm{int}$ and $\theta_\mathrm{scat}$ completely
dominates. If $\theta_\mathrm{scat}$ is only slightly larger than $\theta_\mathrm{int}$, 
then $1 < k < 2$.

\begin{table*}
\begin{minipage}{165mm}
\caption{Kendall's $\tau$-test correlation statistics for $k$-index values and rotation 
         measures $RM$, angular broadening $\theta_\mathrm{scat,\,1GHz}$ from the NE2001 
         model \citep{NE2001}, and H$\alpha$ intensity. Errors are given at 95\% 
         level of significance.}
\label{t:k_vs_rm_ne_ha}
\begin{tabular}{lcccccccccccc}
\hline
\noalign{\smallskip}
Correlation && \multicolumn{3}{c}{$|b|<10\degr$} && \multicolumn{3}{c}{$|b|>10\degr$} && \multicolumn{3}{c}{$|b|>0\degr$} \\
                   \cline{3-5}           \cline{7-9}           \cline{11-13} \noalign{\smallskip}
            && $\tau$ & $N$ & $p$ && $\tau$ & $N$ & $p$ && $\tau$ & $N$ & $p$  \\
   ~~~~~(1) &&    (2) & (3) & (4) &&    (5) & (6) & (7) &&    (8) & (9) & (10) \\
\hline\noalign{\smallskip}
$k$ vs $RM$                          && $0.122\pm0.034$ & 335 & $5\times10^{ -3}$ && $0.058\pm0.015$ & 1992 & $1\times10^{-4}$ &&  $0.095\pm0.014$ & 2327 & $7\times10^{-14}$ \\
$k$ vs $\theta_\mathrm{scat,\,1GHz}$ && $0.188\pm0.033$ & 335 & $3\times10^{ -7}$ && $0.078\pm0.014$ & 1992 & $2\times10^{-7}$ &&  $0.137\pm0.013$ & 2327 & $5\times10^{-23}$ \\
$k$ vs H$\alpha$                     && $0.249\pm0.030$ & 335 & $1\times10^{-11}$ && $0.083\pm0.015$ & 1992 & $4\times10^{-8}$ &&  $0.146\pm0.012$ & 2327 & $9\times10^{-26}$ \\
\hline
\end{tabular}

\smallskip
Note: $N$ is the number of sources, $p$ is the probability that the correlation occurred by chance.
\end{minipage}
\end{table*}

In order to test that the different resolution of 2 and 8~GHz VLBI data does not 
affect the robustness of the derived $k\sim1$ value (Table~\ref{t:gaussians})
and make sure that this result is not due to degradation of angular resolution 
towards lower frequencies, we performed the same analysis for the Galactic plane 
sources using matching resolution at both frequencies. The maximum $(u,v)$ radius 
$r_{uv}^\mathrm{max}$ at 8~GHz was limited to that derived from the corresponding 
2~GHz data sets. The resulting typical value of $r_{uv}^\mathrm{max}$ in the data 
was about 65~M$\lambda$. We took steps to avoid blending effects that lead to the 
innermost portions of the jet component being fitted as part of the core component 
in the 8~GHz data, thereby falsely increasing the fitted sizes of the VLBI core 
component. To implement this, we fit the core component at both frequency bands as 
an elliptical Gaussian, the major axis of which typically aligns with the innermost 
jet direction \citep{2cmPaperIV}, while a minor axis $\theta_\mathrm{min}$ sets 
the transverse width of the outflow in the radio core. The values of $k$-index were 
derived using the VLBI core widths $\theta_\mathrm{min}$ transverse to the jet 
direction. The obtained distribution for $|b|<10\degr$ was fitted with two Gaussians 
with the best-fit peaks at 0.98 and 1.70, very close to the peaks derived from 
the full resolution data sets (Table~\ref{t:gaussians}).

\begin{figure}
\centering
\resizebox{\hsize}{!}{\includegraphics[angle=-90]{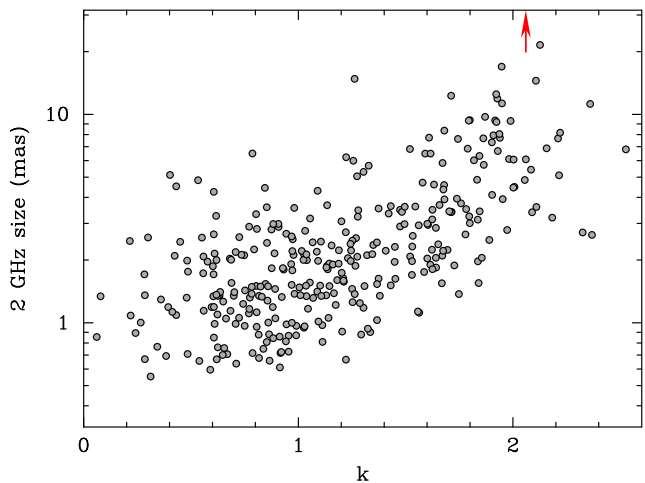}}
\caption{Observed angular core size at 2~GHz as a function of a two-frequency 
         $k$-index for 335 sources within the Galactic plane $|b|<10\degr$. \sgr\ 
         ($\theta_\mathrm{2\,GHz}\approx220$~mas) is beyond the plot limit and 
         indicated by the arrow.
\label{f:k_vs_size}
}
\end{figure}

In Fig.~\ref{f:k_vs_size} we present the expected dependence between the core size
at 2~GHz and $k$-index. It confirms that large observed sizes are indeed found from
strongly scattered sources with high $k$ instead of being intrinsic to the jets.
The distribution of galactic longitude for sources with $k>1.4$ was confirmed 
to be similar to that in Fig.~\ref{f:size_vs_l} (top).

Of the 2327 analyzed sources, 1594 (68\%) have measured redshifts distributed in the
range of $0\lesssim z \lesssim4.7$. We tested a relationship between the $k$-index 
and redshift and found no significant dependence, confirming that the screens which 
dominate in scattering are located in our Galaxy. This is consistent with other studies 
that did not find significant redshift dependence of scatter broadening using VLBI but 
for a much smaller sample of 58 sources \citep{Lazio08}, and using interstellar 
scintillation of 128 sources \citep{Koay12}.

\begin{figure}
\centering
\resizebox{\hsize}{!}{\includegraphics[angle=-90]{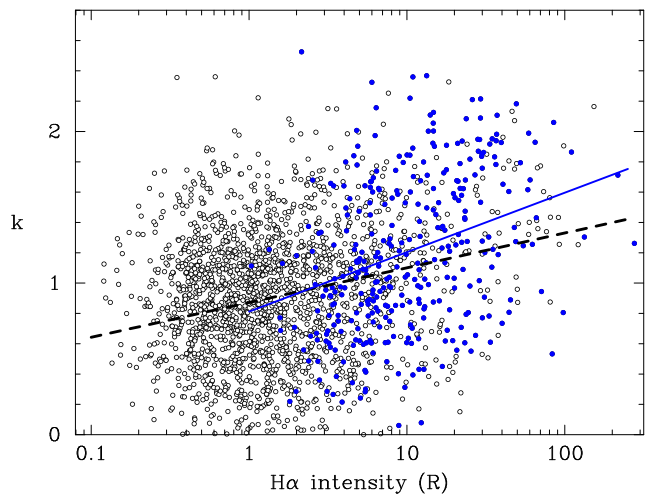}}
\caption{Correlation between the $k$-index for 2327 extragalactic radio sources and
H$\alpha$ intensity in Rayleighs ($1R=10^6$ photons\,cm$^{-2}$\,s$^{-1}$\,sr$^{-1}$) measured
along the corresponding position of the sky. Filled dots represent the Galactic plane sources.
Solid and dashed lines are the least squares fit to Galactic plane and all-sky samples, 
respectively. The fits differ significantly.}
\label{f:k_vs_halpha}
\end{figure}

\begin{table*}
\begin{minipage}{104mm}
\caption{Angular size statistics.}
\label{t:sizes}
\begin{tabular}{r c r c r r r}
\hline
\noalign{\smallskip}
Frequency & $\theta_\mathrm{med}$ ($|b|>10\degr$) & N & $\theta_\mathrm{med}$ ($|b|<10\degr$) & N & $\theta_\mathrm{max}$ & $\theta$ (\sgr) \\
(GHz)     & (mas)&      & (mas)&     & (mas) & (mas)      \\
(1)       &  (2) &  (3) &  (4) & (5) &  (6)  &  (7) \\\hline\noalign{\smallskip}
 2.3      & $1.29\pm0.03$ & 2502 & $1.97\pm0.15$ & 386 & 21.55 & 220  \\
 5.0      & $0.74\pm0.03$ & 1988 & $1.12\pm0.21$ & 120 & 19.16 &  43  \\
 8.4      & $0.37\pm0.01$ & 2606 & $0.44\pm0.04$ & 413 &  6.37 &  14  \\
15.4      & $0.16\pm0.01$ &  464 & $0.23\pm0.05$ &  48 &  1.04 &   4  \\
24.4      & $0.13\pm0.01$ &  289 & $0.15\pm0.03$ & 113 &  0.81 & 1.6  \\
43.1      & $0.07\pm0.01$ &  109 & $0.07\pm0.03$ &  10 &  0.22 & 0.5  \\
\hline
\end{tabular}

\smallskip
Note: the following is presented in the columns:
(1) -- central observing frequency,
(2) -- median angular size for AGN cores outside the Galactic plane and its error at 95\% level of significance,
(3) -- number of sources used to estimate (2),
(4) -- median angular size for AGN cores inside the Galactic plane and its error at 95\% level of significance,
(5) -- number of sources used to estimate (4),
(6) -- all-sky maximum of measured core angular size,
(7) -- angular size of \sgr\ for comparison.
\end{minipage}
\end{table*}

\subsection{Testing connection with rotation measure, free-electron density, and $H\alpha$ Galactic distributions}

Positions of the scatter-broadened sources are expected to match with the sky regions where an
excess of electron density is measured. To test this idea we searched for correlations between 
$k$-index values with
(i)   angular broadening at 1~GHz, $\theta_\mathrm{scat,\,1GHz}$, derived from the NE2001
      Galactic free electron density model, which in turn is based on pulsar observations and also includes 
      scattering properties of extragalactic and other Galactic radio sources \citep{NE2001},
(ii)  absolute value of rotation measures \citep*{Taylor09}, and
(iii) H$\alpha$ emission used as a tracer of ionized gas \citep{Finkbeiner03}.
In all cases, highly significant correlation was established using the non-parametric Kendall's
$\tau$-test (Table~\ref{t:k_vs_rm_ne_ha}). The strongest correlation shown in Fig.~\ref{f:k_vs_halpha}
is detected between the $k$-index derived from our analysis and H$\alpha$ intensity. This is most
probably due to the highest resolution of H$\alpha$ data among the three data sets used. It has
the angular resolution reaching 2\farcm5-square-pixels at $b=0\degr$ from the full-sky composite
H$\alpha$ map \citep{Finkbeiner03}. The observed correlations confirm the extrinsic origin of the 
$k\approx2$ frequency dependence of the AGN core sizes in the Galactic plane.

\subsection{Intrinsic AGN core size: theory predictions and experimental results}
\label{s:jet_theory}

Theoretically, the $r_\mathrm{core}\propto\nu^{-1}$ dependence, where $r_\mathrm{core}$ is the
distance of the apparent jet base (radio core) at a given frequency from the true jet origin,
was predicted by \cite{BK79} in their idealized model of a steady radio jet assuming that
  (i) the jet is conical with a small half-opening angle $\varphi$,
 (ii) the jet is supersonic and free, i.e.\ $\varphi>\mathcal{M}$, where $\mathcal{M}$ is the Mach number,
(iii) the jet has constant velocity,
 (iv) a power-law energy distribution $N(E)\propto E^{-2}$ along the jet,
  (v) there is an approximate equipartition between jet particle and magnetic field energy densities.
The assumption of a conical geometry \citep[see observational evidence by][]{Asada12_M87,Pushkarev14}
leads to $\theta_\mathrm{core}\propto r_\mathrm{core}\propto\nu^{-1}$ dependence.

This is consistent with what we have derived for the majority of the non-scattered sources
(Fig.~\ref{f:k_hists}, Table~\ref{t:gaussians}), and also with the opacity-driven core shift
effect results \citep[e.g.,][]{Lobanov98,Kovalev08_0850,Sullivan_09cs,Sokolovsky_cs,kutkin14}.
Despite the fact that the $k$-index distribution is centered close to 1, its wide 
distribution indicates that departures from the model assumptions are possible in many cases.
This can be caused by, e.g,, pressure and density gradients in the outflow \citep{Lobanov98} 
or non-conical jet geometry. Errors of the derived $k$-index values additionally widen the 
distribution. We have analyzed a sub-sample of 344 sources out of the Galactic plane which have 
rich multi-frequency data of the angular size measurements covering 4, 5 or 6 frequencies. 
From this multi-frequency data, we found that in 25\% of the 344 sources, the k-index values 
do significantly deviate from the value of 1. Nevertheless, based on the large number statistics, 
it is likely that the k-index distribution does indeed peak at a value close to 1.

\section{Scattering properties of \sgr}

In Fig.~\ref{f:size_vs_b} (star symbol), we also show \sgr, a compact radio 
source in the center of the Galaxy. The angular size of the object was 
calculated using a size-wavelength fit $1.0324\,\lambda^{2.0598}$ derived by 
\citet{Zensus07}, giving about 14~mas at 8~GHz and 220~mas at 2~GHz that are 
larger by a factor of about 2 and 10, respectively, than the maximum apparent 
size of an AGN core in our sample. Angular size statistics is given in Table~\ref{t:sizes},
where the errors of the median sizes were estimated using the randomization technique
described in Sect.~\ref{s:app_core_sizes}.
The size of \sgr at 8 and 15~GHz calculated from the fit derived by \citet{Zensus07} 
is consistent with later measurements from other studies \citep[e.g.,][]{Bower14}.

Why is the angular size of \sgr\ so unusual? We discuss below two possibilities.
First, the peculiarity of \sgr\ might be an observing bias if we miss sources 
with comparably large angular sizes as a result of lack of data for imaging 
technique that requires at least four stations to closure amplitudes. Indeed, 
the correlated flux density of a heavily resolved source quickly drops with 
increasing baseline projection \citep[e.g., Fig.~2 in][]{gwinn14} and attains 
the typical VCS detection limit of about 50~mJy \citep{vcs5}. To 
test this scenario, we estimated the minimum total flux density level 
$S_\mathrm{min}^\mathrm{tot}$, below which we can potentially miss large-sized 
sources. Considering a circular Gaussian source model with a size of \sgr\ at 
8~GHz, 14.4~mas, we found that the shortest VLBA baseline projections formed 
by four inner stations reach the VCS detection limit of $\approx50$~mJy if 
$S_\mathrm{min}^\mathrm{tot}=$~100-150~mJy, depending on $uv$-plane coverage.
Remember that $S^\mathrm{tot}>1$~Jy for \sgr\ at 8 GHz. Note that in the RDV 
sessions, in addition to the VLBA stations, a number of other antennas participated 
(normally 6--8 additional radio telescopes) forming an array with shorter baseline 
projections. This implies that the level of $S_\mathrm{min}^\mathrm{tot}$ in the 
RDV data sets is even lower than that derived from the VCS data. We conclude that 
it is unlikely that our sample, which is flux density complete down to about 200~mJy 
at 8.6~GHz above declination $-30\degr$ \citep{vcs5}, misses sources with large observed
angular sizes similar to that of \sgr\ at the corresponding observing frequency. 

\begin{figure}
\centering
\resizebox{\hsize}{!}{\includegraphics[angle=-90]{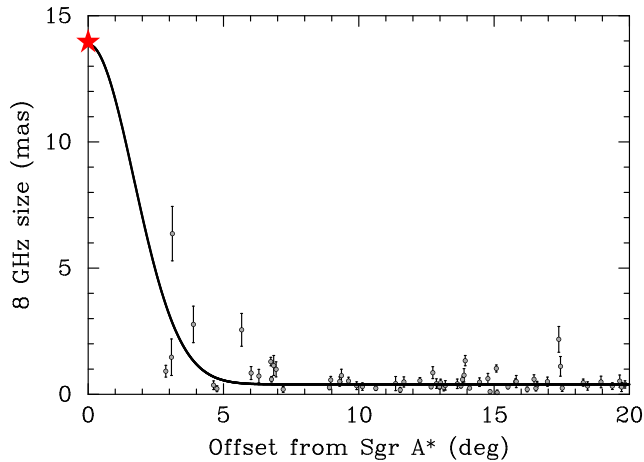}}
\caption{
Observed angular size of AGN cores at 8~GHz as a function of angular separation from \sgr\ shown
by a star symbol, with a clear negative dependence fitted by a Gaussian with the FWHM
$\sim$4\degr and background level $\sim$0.4~mas, suggesting the hyperstrong scattering
screen in the immediate vicinity of \sgr.
\label{f:fig6_fit}
}
\end{figure}

We miss sources that are either a few times larger in size than \sgr\ like the 
extragalactic radio source NGC~6334B --- the most strongly scattered object known, with 
an angular size of 3\arcsec\ at 20~cm wavelength \citep{Trotter98,Moran90}, or 
considerably weaker sources. In principle, an obvious reason for the seeming uniqueness 
of \sgr\ could be an insufficient source density of sky coverage (see the GC region 
in Fig.~\ref{f:fig6_fit}) and deepness of the all-sky sample being analyzed.

The second possibility is that the scattering strength of the screen towards \sgr\ could be 
extremely strong making it the most heavily scatter-broadened source at centimeter wavelengths.
This suggests that a turbulent scattering screen with substantially enhanced free-electron 
density is located in the immediate vicinity of \sgr. Scattering strength peaking at the GC 
decreases with the angular distance from it, and is fitted by a Gaussian with FWHM of about 
4\degr (Fig.~\ref{f:fig6_fit}), providing an estimate of a size for the intermediate-strength
scattering region. Observing radio sources in the GC region, \citet{Roy13} found that scattering 
sizes decrease linearly with increasing angular distance from the GC up to about 1\degr. The 
sources in our sample are not very close to the GC. One of the nearest object, J1744$-$3116, 
observed at 8~GHz is 3\degr away from \sgr\ and has a scatter-broadened source size that 
is smaller by a factor of $\sim$2. The only source in the GC region with an angular size 
comparable to \sgr\ measured at 8.7 and 15.4~GHz by \citet{Bower14}, is the recently discovered 
magnetar gravitationally bound to \sgr\ about $3\arcsec$ away, or at linear separation of 
$\approx0.07-2$~pc \citep{Rea13}. Assuming a single thin-screen model, \citet{Bower14} used 
the combination of temporal smearing \citep{Spitler14} and angular broadening of the magnetar 
emission as a powerful tool for assessing the distance and concluded that the screen is located 
roughly 6~kpc from \sgr towards the Sun. On the other hand, it seems very unlikely that the line 
of sight to the cloud of the ionized medium with the extremely strong scattering strength 
in the Galaxy passes by chance exactly through the GC.

We can place some constraints on the scattering screen media using relations of a thin-screen 
model. First, as it is shown by \citet{Vandenberg76,Lazio08}, the scattered size 
$\theta_\mathrm{scat}\propto n_eD_{LS}$, where $D_{LS}$ is the distance from a source to the 
scattering screen. Therefore, a close (0.1--1~pc from \sgr) screen must be a factor of 
$10^3-10^4$ more dense than the one located roughly halfway \citep{Bower14} to broaden a 
source to the same angular size. Secondly, $\theta_\mathrm{scat}\propto \Delta n_e/\sqrt{a}$ 
\citep{Lyne12}, where $\Delta n_e$ is the fluctuation in electron density on a linear scale 
$a$. Therefore, an alternative scenario for the close screen is that it is more turbulent, 
with a smaller inner length-scale of turbulence \citep*{Armstrong95} estimated 
to be about 300~km in size \citep{gwinn14}. It is also possible that both the increased density and 
smaller inner scale work jointly resulting in the very strong scattering. However, it should 
be noted that a close (to \sgr) screen scenario is strongly inconsistent with the reported 
detailed properties of the magnetar \citep[e.g.,][]{Bower14,Spitler14}.

\section{Summary}

We have found a highly significant negative correlation between the angular size of about 3000 
compact extragalactic radio sources observed with VLBI at 2, 5, and 8~GHz and the absolute value 
of galactic latitude. At higher frequency, 15~GHz, the correlation is significant only for sources 
within the Galactic plane. At both 24 and 43~GHz, where scattering becomes considerably weaker,
no significant correlation is present.

The k-index (with larger values, $\sim$2, being associated with scattering) is found to 
correlate with rotation measure, free electron density, and H$\alpha$ emission distributions 
over the Galaxy. The strongest correlation is established between the $k$ index and H$\alpha$ 
intensity that directly traces ionized gas with high resolution. No significant connection 
is established between the $k$-index and redshift, strongly suggesting that the screens which 
dominate in scattering are situated in the Galaxy. 

Only about one third of AGN viewed through the Galactic plane ($|b|<10\degr$) by VLBI show 
significant angular broadening caused by interstellar scattering. The positional distribution 
of these objects forms three major regions on the sky where scattering is essential, marking 
the Galactic center, the Cygnus region, and active star-forming regions in the Perseus and 
Local arms at galactic longitudes $l\approx220\degr-260\degr$ known as the Fitzgerald 
window \citep{Vazquez08}. At higher galactic latitudes ($|b|>10\degr$) we have found no 
positional clustering of scattered sources. The fraction of sources with $k>1.8$ decreases 
to 4\% for sources at $|b|>10\degr$. This suggests that significant angular broadening should 
not be expected outside of the Galactic plane within the \textit{RadioAstron} \citep{RA2013} 
space VLBI survey of AGN. However, the extreme angular resolution of \textit{RadioAstron} 
at 18, 6 and 1.3~cm could allow observers to detect lower magnitudes of scatter broadening.

The angular size for the majority of non-scattered AGN cores scales approximately as $\nu^{-1}$ 
following the model prediction of a conical jet with synchrotron self-absorption and equipartition 
between the particles and magnetic field energy density \citep{BK79,Koenigl81}. We note that the 
distribution around $k=1$ is wide which hints at many cases with departures from this typical picture.

The angular size of \sgr\ is the largest compared to that of thousands of AGN all over the sky 
observed from 2 to 43~GHz, more than a factor of 10 larger than the maximum AGN core size 
found at 2~GHz. This suggests that \sgr\ is scattered by a compact hyper-turbulent screen 
encompassing the source itself. However, this conclusion is in tension with convincing results of 
\citet{Bower14,Spitler14,Wucknitz14} for the magnetar near \sgr. This points to the scattering 
properties towards the Galactic center being more complex than that described in current models.

The next significant step in studying scattering is expected from an analysis of AGN and pulsars 
at extreme angular resolutions being performed by the Space VLB interferometer \textit{RadioAstron} 
and comparison with predictions of a scattering substructure \citep{jg15,popov15}.

\section*{Acknowledgments}
We thank C.~Gwinn, M.~Johnson, G.~Bower, A.~Deller, D.~Jauncey, O.~Wucknitz, I.~Pashchenko, 
E.~Kravchenko, L.~Petrov, M.~Popov, K.~Sokolovsky for discussions and comments. 
We thank the anonymous referee for useful comments which helped to improve the manuscript.
We deeply thank the teams referred to in Sect.~\ref{s:data_meas} for making their fully 
calibrated VLBI FITS data publicly available and Leonid Petrov for maintaining at the 
Astrogeo Center\footnote{\url{http://astrogeo.org/vlbi_images/}} the database of brightness 
distributions, correlated flux densities and images of compact radio sources produced with 
VLBI. This research has made use of data from the MOJAVE database that is maintained by 
the MOJAVE team \citep{MOJAVE_maps}. This study makes use of 43~GHz VLBA data from the 
VLBA-BU Blazar Monitoring Program, funded by NASA through the Fermi Guest Investigator 
Program. The National Radio Astronomy Observatory is a facility of the National Science 
Foundation operated by Associated Universities, Inc. This study was supported in part by 
the Russian Foundation for Basic Research grant 13-02-12103. This research has made use 
of NASA's Astrophysics Data System.

\bibliographystyle{mn2e}
\bibliography{pushkarev}

\label{lastpage}

\end{document}